\documentclass[superscriptaddress,aps,prl,twocolumn]{revtex4}
\usepackage{bm}
\usepackage{ulem}
\usepackage{epsfig}
\usepackage{graphicx}
\usepackage{amssymb,amsmath,amsbsy,amsgen,amsfonts}    
\usepackage{dcolumn}
\usepackage{amsthm}
\usepackage{mathrsfs}
\usepackage{latexsym}
\usepackage{array}
\usepackage{color}
\usepackage{amstext}
\allowdisplaybreaks[1]
\usepackage{txfonts}
\usepackage{pifont}
\usepackage{tabularx}

\usepackage{epstopdf} 

\newcommand{\ket}[1]{\left\vert{#1}\right\rangle}

\newcommand{\be}{\begin{equation}}
\newcommand{\ee}{\end{equation}}
\newcommand{\ba}{\begin{array}}
\newcommand{\ea}{\end{array}}
\newcommand{\bqa}{\begin{eqnarray}}
\newcommand{\eqa}{\end{eqnarray}}

\DeclareSymbolFont{symbols}{OMS}{cmsy}{m}{n}

\begin{document}

\title{Quantum random number generation using an on-chip plasmonic beamsplitter}

\author{Jason Francis}
\author{Xia Zhang}
\affiliation{School of Chemistry and Physics, University of KwaZulu-Natal, Durban 4001, South Africa}
\author{\c{S}ahin K. \"Ozdemir}
\affiliation{Department of Engineering Science and Mechanics, Pennsylvania State University, University Park, PA 16802, United States of America}
\author{Mark Tame}
\email{markstame@gmail.com}
\affiliation{School of Chemistry and Physics, University of KwaZulu-Natal, Durban 4001, South Africa}

\date{\today}

\begin{abstract}
We report an experimental realisation of a quantum random number generator using a plasmonic beamsplitter. Free-space single photons are converted into propagating single surface plasmon polaritons on a gold stripe waveguide via a grating. The surface plasmons are then guided to a region where they are scattered into one of two possible outputs. The presence of a plasmonic excitation in a given output determines the value of a random bit generated from the quantum scattering process. Using a stream of single surface plasmons injected into the beamsplitter we achieve a quantum random number generation rate of 2.37 Mbits/s even in the presence of loss. We characterise the quality of the random number sequence generated, finding it to be comparable to sequences from other quantum photonic-based devices. The compact nature of our nanophotonic device makes it suitable for tight integration in on-chip applications, such as in quantum computing and communication schemes.
\end{abstract}


\maketitle

{\it Introduction.---} Plasmonic systems exhibiting quantum effects are currently being explored for their potential in emerging quantum technologies~\cite{Tame13}. Compared to standard dielectric systems traditionally used in photonics, plasmonic systems provide a means to confine light to much smaller scales, far below the diffraction limit~\cite{Takahara97,Takahara09}. In the classical regime, this has opened up a range of applications, including nano-imaging~\cite{Kawata09}, nano-sensing~\cite{Homola99,Anker08}, hybrid electro-optic circuitry~\cite{Ozbay06} and new kinds of photonic materials~\cite{Soukoulis11,Meinzer14}. The high confinement of light is achieved by the interaction of the electromagnetic field with free electrons on the surface of a metal to form a joint state of light and matter -- a surface plasmon polariton (SPP). In the quantum regime, studies have shown the enhancement of the interaction between single SPPs and emitters, such as quantum dots~\cite{Buckley12} and nitrogen vacancy centres~\cite{Aharonovich11}. This enhancement has been used to demonstrate compact single-photon sources~\cite{Akimov07,Kolesov09,Huck11,Cuche11} and design new schemes for single-photon switches~\cite{Chang07,Kolchin11,Chang14}, both of which are important devices for quantum technology~\cite{OBrien05}. Other studies have shown that SPPs maintain well the quantum features of the photons used to excite them, including quantum correlations~\cite{DiMartino12}, quantum interference~\cite{Heeres13,Fakonas14,DiMartino14,Cai14,Fujii14} and entanglement~\cite{Alte02}. 

Despite many promising results for plasmonic systems in the quantum regime, they are well known to be inherently lossy~\cite{Maier07} and as such their application to quantum information processing has not been well developed. Initial studies have shown that entanglement can be generated by using plasmonic waveguides even in the presence of loss, either by using loss as a resource~\cite{Gonz11}, or circumventing it with appropriate methods~\cite{Lee13}. By using particular types of encoded quantum states it has also been shown that one can propagate quantum information over arbitrary distances on lossy plasmonic waveguides~\cite{Hanson15}. Most recently, it has been shown that the sensitivity in plasmonic sensing can be enhanced beyond the classical regime by using quantum resources, even in the presence of loss~\cite{Fan15,Pooser15,Lee16}. It is clear from these studies that despite their significant loss, plasmonic systems are still able to exploit quantum effects and carry out useful quantum tasks. The added benefit is that they can do these tasks at a much smaller scale and footprint than conventional photonics, which makes plasmonics an attractive platform for compact nanophotonic quantum information processing.

In this work we further strengthen the use of plasmonics for quantum information processing. We experimentally demonstrate a quantum random number generator using a plasmonic beamsplitter. Quantum random number generators exploit the inherent randomness that is central to quantum mechanics, which makes them ideal sources of entropy. Using classical methods, true random number generation is hard to accomplish, as the unpredictability of the generation relies on an incomplete knowledge of a given system. On the other hand, in a quantum system true randomness is an essential part of the underlying quantum mechanics. True random numbers are important in many applications in science and technology, including in cryptography, simulation of economic, traffic and agricultural models, and for coordination in computer networks~\cite{Her16}. A device that generates true random numbers using quantum mechanics is therefore an important technological component. 

\begin{figure*}[t]
\centering
\includegraphics[width=17cm]{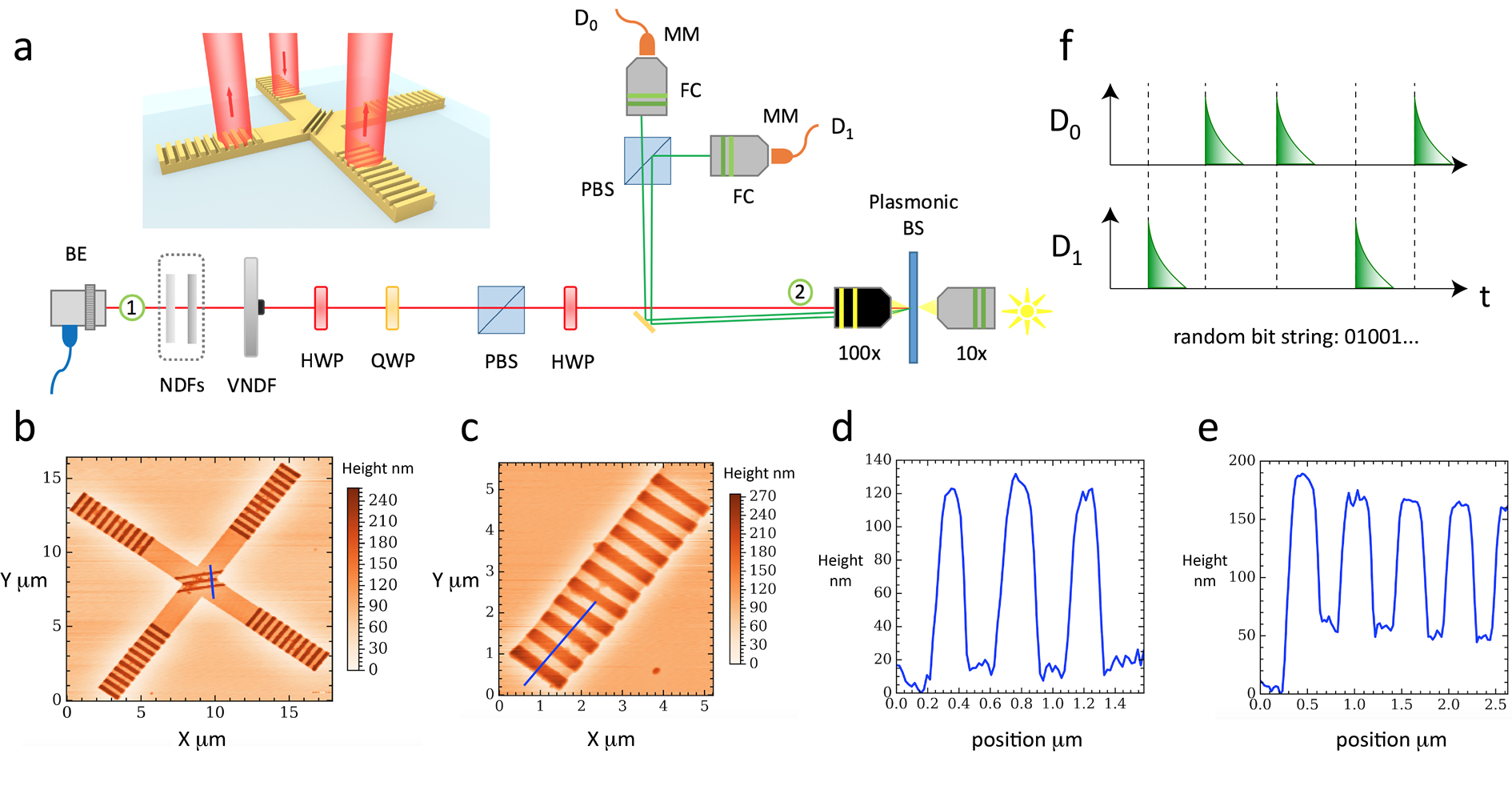}
\caption{Quantum random number generation using a plasmonic beamsplitter. {\bf a}, Microscope stage (inset illustration shows the input and output beam directions for the beamsplitter). The following labels are used: beam expander (BE), neutral density filters (NDFs), variable neutral density filter (VNDF), half-wave plate (HWP), quarter-wave plate (QWP), polarising beamsplitter (PBS), fibre coupler (FC), multimode fibre (MM), beamsplitter (BS) and detector (D). Coherent light is injected into the microscope stage via the BE, and ND filters are used to attenuate it down to the single-photon level before being converted into single surface plasmon polaritons (SPPs). The presence of a single SPP input to the plasmonic beamsplitter is postselected by detection of a photon at either detector D$_0$ or D$_1$, with a random bit being generated. The bit sequence generated from repeated runs originates from the quantum scattering process of single SPPs in the beamsplitter. {\bf b,} Atomic force microscope (AFM) image of the beamsplitter. {\bf c,} AFM image of the bottom left grating. {\bf d,} AFM profile image of the splitting region (zero height set at waveguide level). {\bf e,} AFM profile image of the grating region (zero height set at substrate level). {\bf f,} An illustration of the detector pulses that signal the detection of a single photon at either beamsplitter output, leading to the generation of a random bit sequence (string).
}
\label{fig1} 
\end{figure*}

Previously, single photons have been used as quantum generators of random numbers in bulk setups in various forms, including a branching path generator~\cite{Rarity94,Stefanov00,Jenn00}, time of arrival generator~\cite{Ma05,Stip07,Yu10}, photon counting generator~\cite{Furst10,Ren11,Lopes14} and many others~\cite{Her16}. Most recently on-chip quantum random number generators have been realised using one or more of the above methods~\cite{Sang14,Khan15, Tisa15,Abellan16}. In our experiment, we use the branching path method to generate random numbers quantum mechanically. We excite single SPPs on a gold stripe waveguide and scatter them into one of two possible outputs of a plasmonic beamsplitter. This enables a true random bit to be generated from the quantum scattering process. We then characterise the quality of the random number sequence generated, finding it to be comparable to other methods of photonic-based quantum random number generation~\cite{Her16}. The benefits of a plasmonic splitting device is that it is highly compact and therefore suitable for tight integration in an on-chip setting, where it could be used as a module in a quantum computing or quantum communication task.

{\it Experimental setup.---} The setup used to investigate quantum random number generation using the plasmonic beamsplitter is shown in Fig.~\ref{fig1}a, where a compound microscope is used to excite single SPPs on a plasmonic beamsplitter made from gold, illustrated in the inset of Fig.~\ref{fig1}a. Each arm of the beamsplitter is 2~$\mu$m in width and 70~nm in height. At the ends of the arms and the intersection there is a surface-relief grating of height 90~nm. The end gratings with 11 steps of period 700~nm serve as inputs and outputs for converting photons to SPPs and back again, while the centre grating with 3 steps of period of 500~nm acts as a partial mirror, providing a splitting~\cite{DiMartino14}. A single SPP excited at one input grating propagates along the waveguide and upon reaching the central grating there is a probability that it is transmitted into the forward output and a probability that it is reflected into the perpendicular output. 

This type of scattering process has been studied in much detail in the classical regime~\cite{Lamprecht2001,Ditlbacher2002,Weeber2005,Gonzalez2006} and more recently in the quantum regime~\cite{DiMartino14}. While the aim of the central grating is to act as a partial mirror, transmitting and reflecting part of the incoming excitation, there is also some scattering into the far-field and absorption in the metal. The reflection, transmission and loss coefficients in the classical regime are translated into probability amplitudes in the quantum regime via energy conservation arguments~\cite{Barnett98}. Such a lossy quantum splitting has been studied extensively for photons~\cite{Barnett98} and surface plasmons~\cite{Ballester10} theoretically, and experimentally observed in recent studies of the plasmonic Hong-Ou-Mandel effect~\cite{Heeres13,Fakonas14,DiMartino14,Cai14,Fujii14}. These results allow us to faithfully model the scattering process as a lossy quantum optical beamsplitter. As such, the branching path model for our source of randomness (entropy) is valid~\cite{Her16,Stip15,Stip15a}. The impact of loss on the scattering process and subsequent random number generation will be discussed later.

During propagation, the SPPs are highly confined to the surface of the waveguide, with a perpendicular characteristic length much less than the free-space wavelength~\cite{DiMartino12,DiMartino14}. This confinement, together with the in-plane footprint of $\sim15 \mu$m~$\times 15~\mu$m makes the plasmonic beamsplitter a highly compact device, comparable in size to current state-of-the-art on-chip quantum photonic waveguide couplers~\cite{Bonneau12,Zhang12,Wang15,Abellan16}. Shorter in/out gratings, lead-in propagation distance to the splitting region and waveguide widths are possible in principle~\cite{Lamprecht2001,Ditlbacher2002,Zia2005,Weeber2005,Gonzalez2006,Zenin16}, and would enable further reduction in size if needed for a given application. In addition, while the beamsplitter in our experiment is excited by an external source and the detection is performed off-chip, as in many recent quantum photonic demonstrations~\cite{Bonneau12,Zhang12,Carolan15}, future work on the integration of sources~\cite{Akimov07,Kolesov09,Huck11,Cuche11,Clemmen09} and detectors~\cite{Heeres13,Falk09} would enable a completely standalone device.

The beamsplitter is fabricated on a silica glass substrate of thickness 0.17~mm (refractive index $n=1.5255$) by a combination of electron beam lithography (EBL) and electron beam evaporation (EBE). For the waveguide sections (layer 1, thickness 70~nm), a positive resist is spin coated on the substrate, with EBL used to define the waveguide regions. A lift-off technique is then used, with an adhesion layer of Ti (thickness 2-3~nm) deposited first and then the Au layer using EBE. The gratings (layer 2, thickness 90~nm) are formed similarly to the waveguides using alignment marks to match up layer 1 and 2. A three-dimensional image of the final beamsplitter structure is then obtained using an atomic force microscope (NT-MDT Smena), as shown in Figs.~\ref{fig1}b-e.

To excite single SPPs in the beamsplitter a coherent light source in the form of a $\lambda=780$~nm continuous-wave laser operating above the lasing threshold at 35~mA is used (Thorlabs LPS-785-FC). The light is injected into the microscope stage via a beam expander (BE). The collimated beam runs through a combination of two fixed and one variable neutral density (ND) filters, which are used to attenuate the coherent light down to the single-photon regime. The prepared source of light can be described by a weak coherent state 
$\ket{\alpha}=e^{-|\alpha|^2/2}\sum_n \frac{\alpha^{n}}{\sqrt{n!}} \ket{n}$, with photon number distribution $p_n=e^{-|\alpha|^2}\frac{|\alpha|^{2n}}{n!}$ and $|\alpha|^2=\langle \hat{n} \rangle \ll 1$ as the mean excitation number. In this regime, the two dominant components are $p_0$ and $p_1$. Using single-photon avalanche diode (SPAD) detectors (Excelitas SPCM-AQRH-15) for each output of the beamsplitter, the vacuum component $p_0$ is removed by postselection, {\it i.e.} when either detector clicks the quantum state injected into the input grating of the beamsplitter was a single photon~\cite{Jenn00}.

To clean up the polarisation of the input state, a polarising beam splitter (PBS) is used to select horizontally polarised photons, while a preceding quarter-wave plate (QWP) and half-wave plate (HWP) are used to control the incident polarisation. The beam is focused onto one input grating of the plasmonic beamsplitter (diffraction-limited spot) using a 100x microscope objective and its polarisation is adjusted using an additional HWP to maximise the in-coupling efficiency for conversion of photons to SPPs~\cite{DiMartino12}. The two output modes of the beamsplitter are collected by the same objective, and are then picked off by a knife-edge mirror and directed onto a PBS. This PBS separates the two orthogonally polarised outputs which are then coupled into multimode fibres that lead to separate SPAD detectors. The detectors are each connected to a channel of a Picoquant TimeHarp 260, which in turn is connected to a PC for data acquisition. In time-tagging mode the TimeHarp records the detector at which a photon arrives and its arrival time to a resolution of $25$~ps. 

In order to excite single SPPs we require that the average photon number $n$ is much less than one per coherence time of the source~\cite{Jenn00}. The coherence time is given by $\tau=\sqrt{2 {\rm ln}2}/\pi \delta \nu$, where $\delta \nu$ is the frequency bandwidth of the laser~\cite{Saleh07}. Using $\delta \nu = 9.74$~THz we have $\tau=3.85 \times 10^{-14}$~s. Thus the rate of photons injected into the plasmonic beamsplitter, $R$, must be much less than $1/\tau=2.60 \times 10^{13}$~s$^{-1}$. Due to the low initial intensity required, the input rate in our setup was calculated by measuring the power at position 1 in Fig.~\ref{fig1}a and then measuring the transmission efficiency of all the optics between position 1 and 2. This yielded a transmission factor of $\eta=2.78 \times 10^{-6}$. With the power at position 1 as $P_1=1.23$~mW we obtain an input power of $P_{in}=\eta P_1=3.77 \times 10^{-9}$~W into the microscope objective. Dividing this by the average photon energy gives $R\simeq \frac{\lambda P_{in}}{hc}=1.47 \times 10^{10}$~s$^{-1}$. While this is clearly much less than $1/\tau$, an additional constraint comes from the SPAD detector dead time $\tau_d=24$~ns~\cite{Her16,Stip15}. If a detector detects a photon it will not detect another within this time. Thus for 2 photons arriving within $\tau_d$ only one will be detected. This leads to an excess of substrings 01 and 10, as the detectors miss the substrings 00 and 11 at random places~\cite{Stip15}. The input power is therefore set using a variable ND filter so that the detected rate in each detector is $1.2 \times 10^6$~s$^{-1}$~$\ll 1/\tau_d=4.2 \times 10^{7}$~s$^{-1}$. 

Losses occur in the setup due to scattering (during interconversion of photons to SPPs at the gratings and the SPP splitting) and absorption (during SPP propagation). These have the same effect on the quantum properties of the excitations and can be combined~\cite{Tame08,Ballester10,DiMartino12}. At the input stage, which includes the grating interconversion efficiency ($\sim 12 \%$ at $780$~nm) and propagation to the splitting region ($e^{-\ell/L_p}\simeq 59 \%$ using waveguide length $\ell=4.5 \mu$m and decay length $L_p=8.5 \pm 0.4~\mu$m), the losses reduce the overall amplitude of the coherent state before it arrives at the splitting region. This reduces the component $p_1$ relative to $p_0$, which therefore reduces the total number of post-selected single SPPs that are input to the splitter. The reduction can be compensated by simply increasing the initial amplitude of the coherent state. At the output stage, which includes the splitting, propagation from the splitting region and grating interconversion (with symmetrical loss in the two outputs), losses remove an equal number of single excitations in the two arms. This leads to a balanced reduction in the number of 0's and 1's measured. The reduction can again be compensated by increasing the initial amplitude of the coherent state. 

In general, when compensating for loss by increasing the initial amplitude of the coherent state it is important to ensure that two factors are taken into consideration. The first is that the beamsplitter must operate in the single-excitation regime and so the average excitation number arriving at the splitting region should be much less than one per coherence time of the coherent state. This limits the amount by which the initial amplitude can be increased. The second factor is that any bias in the random numbers generated should be minimized and so the detected rate from the outputs must be less than the reciprocal of the detector dead time. This also limits the amount by which the initial amplitude can be increased. Both factors depend on the geometry of the beamsplitter (gratings, arm length and splitting efficiency) and the type of detector used. By setting the input power $P_{in}$ in our experiment to the value specified, we have fixed the initial amplitude of the input state in order to satisfy both factors, as detailed above.

At the measurement stage the detection of photons from the plasmonic beamsplitter outputs produce signals in the form of pulses from the SPADs, as illustrated in Fig.~\ref{fig1}f. These signals are sent to the TimeHarp which records the arrival times of the pulses from both channels and from this we extract a random binary sequence. The sequence is constructed by denoting a pulse from detector 0 as the binary digit 0 and a pulse from detector 1 as the binary digit 1. A total of 82,604,923 binary digits (bits) were acquired over a 34~s period, corresponding to a rate of $2.43$~Mbits/s.

\begin{figure}[t]
\centering
\includegraphics[width=7.5cm]{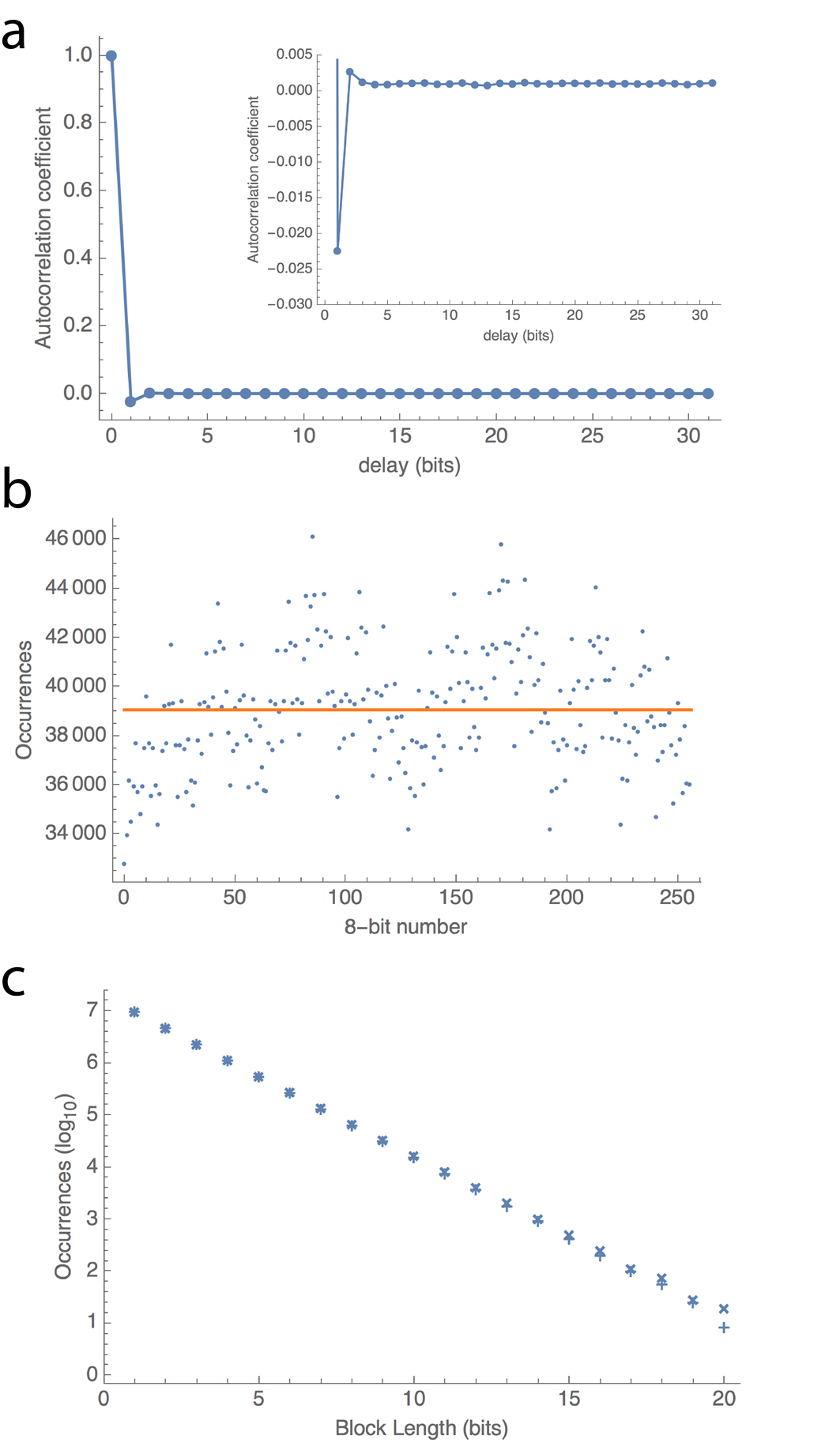}
\caption{Characterisation of generated random number sequence. {\bf a}, Autocorrelation coefficient. Inset shows zoomed region to highlight residual correlation. {\bf b}, Distribution of 8-bit blocks (bytes). The horizontal line represents the average. {\bf c}, Distribution of run lengths for blocks of ones ($\times$) and zeros ($+$).}
\label{fig2} 
\end{figure}

{\it Characterisation of random number sequence.---} We characterise our random bit sequence by applying a number of standardised tests~\cite{Jenn00,Her16}. These tests are employed to determine if the generated sequence exhibits characteristics typical of a true random bit sequence. In what follows we describe the tests performed on a sample of $80$~Mbits and present the results:

(i) Autocorrelation: We first calculate the correlation coefficients~\cite{Edwards76} of the bit sequence with itself up to a 31-bit delay in order to examine any short-ranged correlations and periodicity. The coefficients lie in the interval [-1, 1]. A value of 1 and -1 indicate correlation and anti-correlation respectively, while a value of 0 suggests uncorrelated bits. Fig.~\ref{fig2}a is a plot of the obtained coefficients, and shows only a small negative correlation between adjacent bits, after which the correlation remains consistently close to zero. The inset shows a zoomed region that omits the first data point in order to emphasise the lower lying points. The negative correlation can be seen, as well as a small positive correlation for delays greater than one bit. The main sources of these residual correlations in a beamsplitter scenario can be attributed to a combination of detector dark counts, deadtime and afterpulsing in the SPADs~\cite{Stip15,Stip15a}.

(ii) Uniform distribution of $n$-bit blocks: For a true random bit sequence, all possible $n$-bit combinations should be equally probable. We consider single bits and 8-bit blocks. In the case of single bits we should have an equal number of zeros and ones for a large enough sample. For 8-bit blocks that are converted to unsigned integers, we expect a uniform distribution of occurrences over the domain $[0, 255]$. Furthermore, the average must approach the value of 127.5. Using our 80~Mbit sample we find the proportion of ones as 0.5023 and zeros as 0.4977, showing a small bias towards one. The distribution of 8-bit integers is shown in Fig.~\ref{fig2}b. The average of the $10 \times 10^6$ integers is 128.13, which is slightly larger than expected due to the small bias and residual correlations in the sequence.

(iii) Distribution of run lengths: A run is a continuous string of zeros or ones. If finding a zero or a one are equally likely then the probability of finding a run of $n$-bits is proportional to $2^{-n}$. Fig.~\ref{fig2}c shows the  run length distributions for both zeros and ones. Fitting a straight line to the first twenty points gives a gradient of ($-0.312 \pm 0.001$) for runs of zeros, and $(-0.301 \pm 0.002)$ for runs of ones. The ideal values should be $-{\rm log}_{10}2\simeq-0.301$. The discrepancy in the case of runs of zeros is due to the smaller probability of obtaining a zero as a result of bias. The smaller probability results in a higher likelihood of obtaining a run of ones than a run of zeros of the same length. This in turn contributes to the small periodic character of the 8-bit integer distribution shown in Fig.~\ref{fig2}b. 

(iv) The entropy of an $n$-bit string: This is defined as $-\sum_ip_i {\rm log}_2p_i$, where $i \in \{0,1\}^n$ and $p_i$ is the probability of obtaining $i$. The entropy can be used as a measure of irregularity. A perfectly random source of $n$-bit strings should have $n$ bits of entropy. A value of $7.99726$ bits was calculated for 8-bit strings using the probabilities obtained from the distribution in Fig.~\ref{fig2}b. 

(v) Estimation of $\pi$: As a more functional check we estimated the value of $\pi$ using a Monte Carlo method. Here the area of a circle of radius $r$ divided by the area of a square with sides of length $2r$ is equal to $\pi/4$. Thus, by populating a quadrant of the square with enough randomly placed points and finding the ratio of points within the quarter-circle to the total, we arrive at an estimate of 3.13366 for $\pi$. 

\begin{table}[t]
\centering
\begin{center}
\noindent\begin{tabular*}{\columnwidth}{@{\extracolsep{\stretch{1}}}*{4}{r}@{}}
  \hline
  & Mean~ & Entropy~ & $\pi$~~~~~ \\
  \hline
  \hline
QRNG & 127.49 & 7.999981 & 3.13227 \\
QRNG~\cite{Jenn00} & 127.50 & 7.999965 & 3.14017 \\
PRNG & 127.50 & 7.999982 & 3.13252 \\
\hline            
\end{tabular*}
  \end{center}    
\caption{Summary of results from tests applied to the post-processed and PRNG sequences.}
\label{tab:test} 
\end{table}

Due to a slight asymmetry of the beamsplitter and functionality of the SPADs, the bit strings produced exhibit small short-ranged correlations amongst bits, as shown in Fig.~\ref{fig2}a, and a bias in the form of non-uniform single-bit and 8-bit distributions, as shown in Figs.~\ref{fig2}b and c. To mitigate these effects we employ a randomness extractor to the bit sequence~\cite{Her16}. The extractor applied here is an extension of the von Neumann scheme~\cite{vonNeumann51,Peres92} and was chosen for its simplicity and non-use of a random seed. The algorithm proposed by von Neumann can be applied to a biased generator of independent bits. Such a generator would produce a $0$ and a $1$ with respective probabilities of $p$ and $q$ with $p \neq q$. Since the bits are independent, the bit-pairs $01$ and $10$ occur with an equal probability of $pq$. Thus, occurrences of these pairs can serve as an unbiased source of random bits. This is done by assigning a $0$ bit when a $01$ pair is produced, and a $1$ for $10$ pairs. A biased input will therefore be reduced to a fraction $pq$ of its original length. An extension of this scheme described in Ref.~\cite{Peres92} overcomes the reduction by producing further biased sequences from the original sequence, to which the von Neumann algorithm is then applied. This procedure is repeated recursively with each output being concatenated to the previous. Before applying the recursive algorithm we require that adjacent pairs of bits be uncorrelated. From the autocorrelation plot in Fig.~\ref{fig2}a it is clear that a fraction of adjacent bits are anti-correlated. To minimise this correlation we first shuffle the sequence and then apply the recursive von Neumann algorithm. Applying the algorithm to 32 sequences of 2.4~Mbits yielded an average output of $2.3682 \pm 0.0003$~Mbits, leading to a marginally reduced generation rate of 2.37~Mbits/s.
\begin{figure}[t]
\centering
\includegraphics[width=7.5cm]{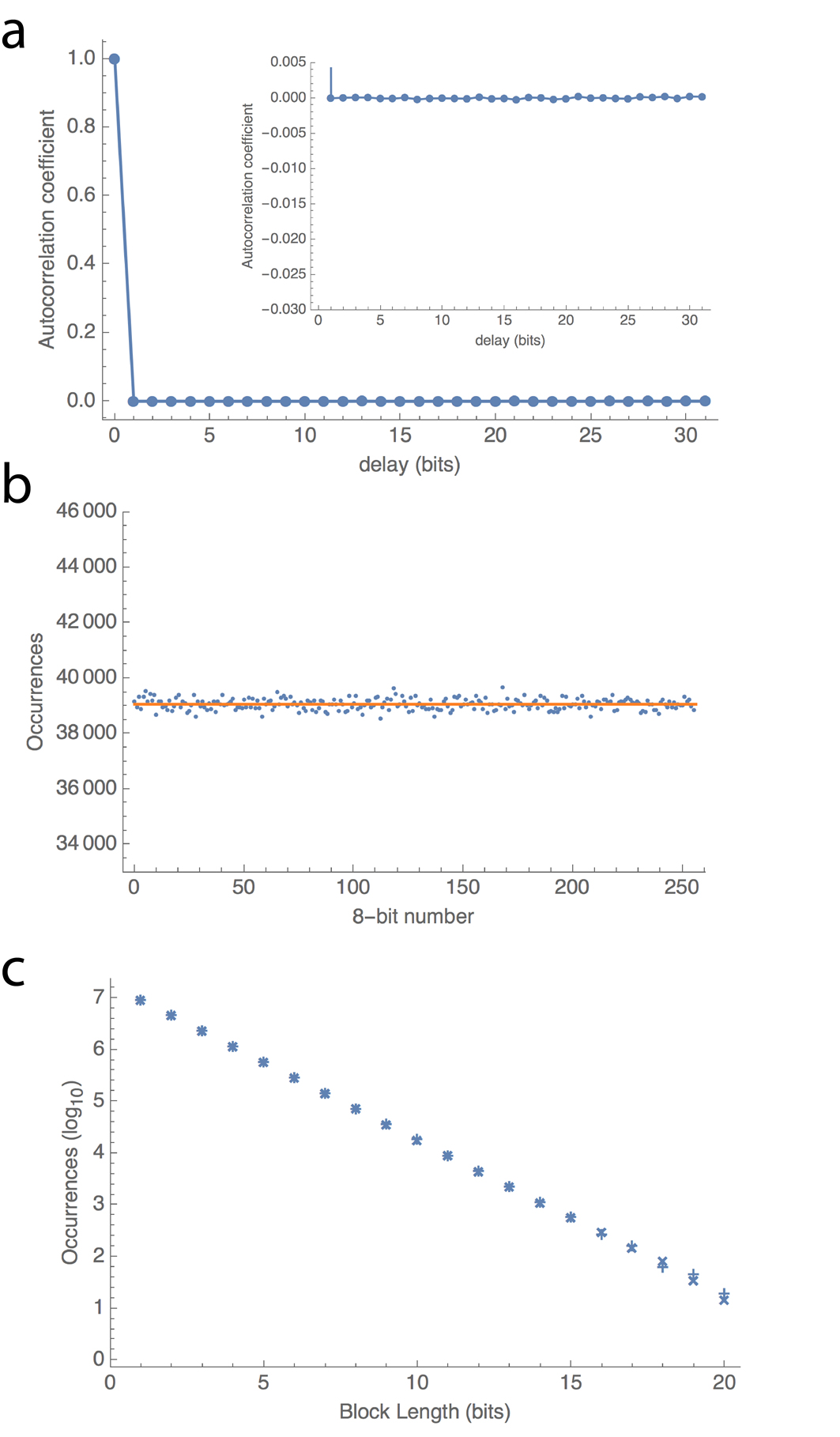}
\caption{Characterisation of post-processed quantum random number sequence (QRNG). {\bf a}, Autocorrelation coefficient. Inset shows zoomed region to highlight residual correlation. {\bf b}, Distribution of 8-bit blocks (bytes). The horizontal line represents the average. {\bf c}, Distribution of run lengths for blocks of ones ($\times$) and zeros ($+$).}
\label{fig3} 
\end{figure}

The tests described above are then applied to the post-processed sequence, which we denote as QRNG. For comparison, a sequence of equal length obtained from the Marsaglia CD-ROM~\cite{Marsaglia95} was also tested. The sequence was generated using a combination of pseudorandom number generators. We denote this sequence as PRNG. A summary of the results are given in Tab.~\ref{tab:test}, where we have also included the results from the quantum photonic implementation of Ref.~\cite{Jenn00}. The detailed results for the QRNG sequence are shown in Fig.~\ref{fig3}. There is clearly an improvement that can be seen in the results of the tests. The autocorrelation varies by a maximum of $\sim0.0002$ over the bit delay range, the distribution of bytes is markedly improved and the distribution of block lengths of zeros and ones match up for larger block sizes. As a further and more rigorous test we apply the NIST Statistical Test Suite (STS) to the QRNG and PRNG sequences. The NIST test suite is aimed at becoming the first industry standard for testing random numbers and a detailed description of the tests can be found in Ref.~\cite{NIST}. The PRNG sequence of 80~Mbits fails the NIST test suite (Longest Run and Overlapping Template tests are not passed), while the QRNG sequence passes all tests at the $1\%$ significance level, performing well compared to other quantum photonic-based implementations~\cite{Her16,Stip07,Furst10,Sang14,Wei09,Nie14,Stip15}. A summary of the results of the NIST tests is given in Tab.~\ref{tab:NIST}. The Overlapping Template, Linear Complexity, Random Excursions and Random Excursions Variant tests were run on a set of 80 sequences of 1~Mbits in length. The remaining tests ran on 160 sequences of length 500~Kbits.
\begin{table}[t]
\centering
\begin{center}
\noindent\begin{tabular*}{\columnwidth}{@{\extracolsep{\stretch{1}}}*{4}{r}@{}}
  \hline
  Statistical Test & p-value~ & Prop/Thr~~~ & Pass \\
  \hline
  \hline
Frequency		 & $0.546791$ & $156/154$~~~~ & Yes \\
Block Frequency	 & $0.624107$ & $159/154$~~~~ & Yes  \\
Cumulative sums	 & $0.606531$ & $158/154$~~~~ & Yes \\
Runs			 & $0.371101$ & $159/154$~~~~ & Yes \\
Longest Run 	 & $0.284375$ & $159/154$~~~~ & Yes \\
Rank 			 & $0.162606$ & $158/154$~~~~ & Yes \\
FFT				 & $0.947557$ & $157/154$~~~~ & Yes \\
Non Overlapping Template	  & $0.723759$ & $158/154$~~~~ & Yes \\
Overlapping Template 	  & $0.559523$ & $79/76$~~~~ 	& Yes \\
Universal				  & $0.330628$ & $159/154$~~~~ 	& Yes \\
Approximate Entropy 	  & $0.350485$ & $157/154$~~~~ 	& Yes \\
Random Excursions 		  & $0.516893$ & $44/42$~~~~ 	& Yes \\
Random Excursions Variant & $0.054933$ & $45/42$~~~~ 	& Yes \\
Serial 					  & $0.606531$ & $160/154$~~~~ 	& Yes \\
Linear Complexity 		  & $0.392456$ & $79/76$~~~~ 	& Yes \\
\hline            
\end{tabular*}
  \end{center}    
\caption{Summary of results from the NIST tests applied to the post-processed sequence QRNG. The p-value is the probability a perfect random number generator would generate the particular experimental result or a result indicating a less random bit sequence. If p~$>0.01$, the significance level, the bit sequence passes the test. Prop is the proportion of the tested sequences that succeed in passing a test. 
Thr is the theoretical lower-bound of acceptable proportion values~\cite{sys15}.}
\label{tab:NIST} 
\end{table}

{\it Discussion.---} In this work we demonstrated the generation of random numbers using a plasmonic beamsplitter operating in the quantum regime. The presence of a plasmonic excitation in a given output from the beamsplitter determined the value of a random bit generated from a quantum scattering process. Using a stream of single plasmons we achieved a quantum random number generation rate of 2.37 Mbits/s, despite the presence of loss in the waveguides. We characterised the quality of the random number sequence generated and with post-processing found it to be comparable in quality to sequences from other photonic-based devices. Higher generation rates in our setup may be achieved in a number of ways: the use of detectors with reduced deadtimes~\cite{Furst10}, operating several beamsplitters in parallel and including additional degrees of freedom such as the excitation arrival time~\cite{Her16,Stip15}. Improvements in the fabrication of the beamsplitter would also reduce the bias between ones and zeros, and together with detector improvements would remove the need for post-processing which adds an additional resource overhead. The compact nature of the beamsplitter makes it suitable for integration in a variety of on-chip applications, such as in quantum computing and communication schemes. Future work into the integration of high quality sources and detectors on-chip would enable a fully self-contained device for use in a range of applications where true random number generation is required.
\begin{figure}[t]
\centering
\includegraphics[width=8.5cm]{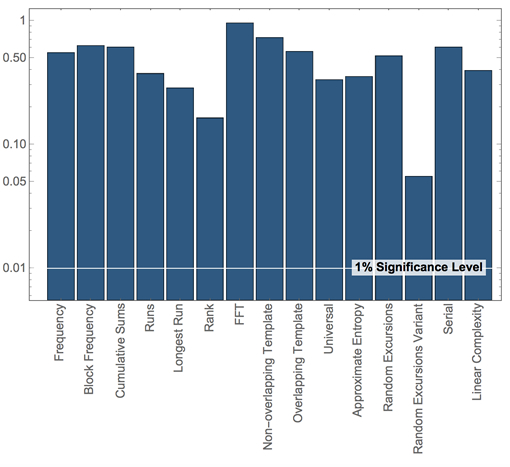}
\caption{Summary of p-value results from the NIST tests applied to the post-processed sequence QRNG. If p~$>0.01$, the significance level, the bit sequence passes the test.}
\label{fig4} 
\end{figure}

{\it Acknowledgments.---} We thank Francesco Petruccione, NTT-AT Japan and NT-MDT. This research was supported by the South African National Research Foundation, the National Laser Centre, the UKZN Nanotechnology Platform and the South African National Institute for Theoretical Physics. S.K.O acknowledges the support of Pennsylvania State University Materials Research Institute (MRI).


\end{document}